# Optimal energy collection with rotational movements constraints in Concentrated Solar Power Plants

José-Miguel **Díaz-Báñez**[a], José-Manuel **Higes-López**[a], Miguel-Angel **Pérez-Cutiño**[a,b,*] and Juan **Valverde**[a,b]

[a]*Department of Applied Mathematics, University of Seville, SPAIN,*

[b]*Virtualmechanics S.L, Seville, SPAIN,*

## ABSTRACT

In Concentrated Solar Power (CSP) plants based on Parabolic Trough Collectors (PTC), the Sun is tracked at discrete time intervals, with each interval representing a movement of the collector system. The act of moving heavy mechanical structures can lead to the development of cracks, bending, and/or displacements of components from their optimal optical positions. This, in turn, diminishes the overall performance of the entire system for energy capture. In this context, we introduce two combinatorial optimization problems to limit the number of tracking steps of the collector and hence the risk of failure incidents and contaminant leaks. On the one hand, the Minimum Tracking Motion (MTM)-Problem aims at detecting the minimum number of movements while maintaining the production within a given range. On the other hand, the Maximal Energy Collection (MEC)-Problem aims to achieve optimal energy production within a predetermined number of movements. Both problems are solved assuming scenarios where the energy collection function contains any number of local maximum/minimum due to optical errors of the elements in the PTC system. The MTM- and MEC-Problems are solved in $O(n)$ time and $O(n^2 m \omega^*)$ time, respectively, being $n$ the number of steps in the energy collection function, $m$ the maximum number of movements of the solar structure, and $\omega^*$ the maximal amplitude angle that the structure can cover. The advantages of the solutions are shown in realistic experiments. While these problems can be solved in polynomial time, we establish the NP-hardness of a slightly modified version of the MEC-Problem. The proposed algorithms are generic and can be adapted to schedule solar tracking in other CSP systems.

## 1. Introduction

The transition from fossil fuels to green technologies has aroused great interest over the past decades. In particular, the European Union has proposed that 45% of the energy it uses come from renewable sources by 2030. In this context, capturing and storing solar energy is a seminal research area, with a special focus on optimization algorithms to reduce operation and maintenance costs (Cirre et al., 2009; Carrizosa et al., 2015; Ashley et al., 2017; Bigerna et al., 2019; Ruiz-Moreno et al., 2021).

Concentrated Solar Power (CSP) plants are an effective alternative to photovoltaic technologies, as it has the capacity of storing the energy captured from the Sun. Parabolic Trough Collectors (PTC) systems are one of the most widespread CSP plants around the globe, including more than 40 plants in Spain alone. PTC systems are composed of a parabolic-shaped surface reflecting the Sun rays to a Heat Collector Element (HCE) located at the focus of the parabola. The HCE carries a Heat Transfer Fluid (HTF), usually synthetic oil, that is used to produce the steam that feeds an industrial process or a power block to produce electricity. The parabolic-shaped mirror surface together with three HCEs forms a Solar Collector Element (SCE), and 4 SCEs are a Solar Collector Assembly (SCA) that are arranged in series to form a full PTC loop. For a full description of the elements in the solar field of PTC plants, the reader is referred to Barcia et al. (2015).

During normal operation of PTC plants, SCAs are automated to follow the Sun (tracking) so that the maximum energy can be collected, see Figure 1. Solar trackers are classified into five types based on their tracking technologies: active tracking, passive tracking, semi-passive tracking, manual tracking, and chronological tracking (Hafez et al.,

---

*Corresponding author

✉ dbanez@us.es (J. Díaz-Báñez); jhiges@us.es (J. Higes-López); m.perez@virtualmech.com (M. Pérez-Cutiño); jvalverde@us.es (J. Valverde)

ORCID(s): 0000-0002-4031-4309 (J. Díaz-Báñez); 0000-0003-1240-3387 (J. Higes-López); 0000-0002-8841-2565 (M. Pérez-Cutiño); 0000-0002-8065-800X (J. Valverde)

---

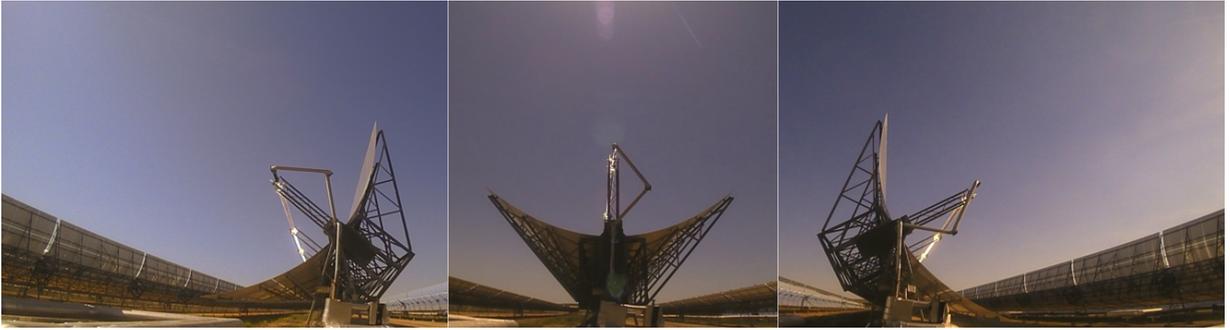

**Figure 1:** Solar tracking example in a real PTC plant: REPA to connect fixed to moving HTF carrying pipes.

2018). However, independently on the installed tracking system, SCAs move at discrete time intervals, producing mechanical stress over the structure at each step. Eventually, the stress results in anomalies in the components of the plant, hindering its overall performance.

The SCA is a large, high-precision optical structure which is tracking/rotated by an hydraulic motor. The structure cost is very important, therefore, it is designed with the minimum amount of material to guarantee the optical precision. Therefore, it is not an over-stiff structure and deforms during operation. The rotation is done step-by-step by the motor and to do so, it has to overcome all rotation resistances exerted by the structure, namely: self-weight, friction at supports, connections (REPA: Rotation Expansion Performing Assembly, see Figure 1) to fixed pipes carrying the HTF from/to the Solar Field, etc. All these elements wear with the pass of time increasing the resistance of the structure to rotate. At certain conditions, the mechanical diode breaks (usually the weakest point being the REPA) and the structure must be put out of service. Actually, chances to detect this incident are very low, incurring most of the time in HTF leakage to ambient, with the subsequent ground contamination and risk of accidents with personnel and other equipment nearby. This is the most severe incident that can happen in a PTC plant and it is the aim of operation and maintenance personnel to avoid these type of issues. Different approaches are being studied in the industry to avoid such kind of incident during operation (Eckhard, 2022; Schiricke, 2019). On the other hand, this problem must be tackled from design phase and this is exactly where the core of the paper is directed: one way to diminish the risk is to limit the number of steps of motion on the collector, which are the precise instants where risk of breakage happens due to the collector rotation against the resistance torques. From an industrial perspective, significant financial resources are allocated to maintenance in order to avoid leak-contaminant events.

When the operating conditions are optimal, a perfect tracking of the Sun results in maximal energy collection. In this context, the optimal operating conditions include: no errors on the tracking system, the SCA is built and installed without errors, the PTC structure is infinitely rigid and does not deform, and the weather conditions are constant throughout the day. Considering this scenario, the ray incidence over the HCE for different SCA and solar angles is expected to be a unimodal function; see Figure 2 for a theoretical example. However, the shape of the function can change due to several factors, such as installation errors of some components of the SCA, structural deformation due to wear and increased torque resistance of the SCA to rotate, cracks/dirt in the mirror surface, HCE bending and vertical/horizontal displacements due to mechanical stress, among others. For multi-modal functions, optimizing the energy collection while reducing the rotations of the SCA is a complex and interesting problem. To the best of our knowledge, this problem has not been formally defined or solved in the existing literature.

## 1.1. Contributions

This paper introduces discrete optimization problems inspired by efficient solar tracking systems when optical or mechanical anomalies are affecting the rays concentration over the HCE, which is the realistic situation when the system is put into operation. Moreover, we consider the movements of the SCA as one of the variables in our optimization scheme; hence our solution aligns with the reduction of the stress in the components of PTC plants (minimizing or limiting the number of movements and therefore minimizing the risk of leak/failure incidents). Our algorithms are based in techniques of discrete optimization, ensuring optimal solutions in pseudo-polynomial time. For this initial study, we require that the weather conditions are constant throughout the day. Although we use as objective function the Sun rays incidence over the HCE, our solution is general and can be used with any variable involving power generation in CSP plants as long as its evolution is known beforehand; for instance, with forecasting methods (Wang and Gao, 2022). In addition, we prove that a close version of one of the problems is NP-Hard, which reinforces the importance of the provided solution.

Our main contributions can be summarized as follows:

- We present efficient algorithms for two optimization problems associated with the design of an optimal schedule in solar tracking. Our proposal considers tracking in scenarios where the objective function contains any number of local maximum/minimum due to error in elements of the PTC system.

- To the best of our knowledge, this paper is the first considering the optimization of the solar tracking while reducing the movements of the SCA in a PTC plant. The advantages of our proposal are demonstrated with realistic experiments.

- We prove the NP-Hardness of a variant of the problem.

## 1.2. Overview

The rest of the paper is organized as follows: Section 2 summarizes the existing literature for solar tracking and maximal energy collection; Section 3 provides the necessary background and the formal definitions of the considered optimization problems; Sections 4 and 5 detail our methodologies for addressing the presented problems. In Section 6, empirical evidence of the advantages of the proposed algorithms is provided through simulation, while Section 7 addresses the NP-hardness of our problem when subjected to a minor modification. Finally, Section 8 summarizes the results and proposes future work of this research.

## 2. Related work

From an industrial perspective, reducing the movements of the SCA is of vital importance to reduce failures in the REPAs of the solar field.To reduce mechanical stress one strategy involves decreasing the movements of the tracking system. Although novel innovative designs of REPAs are tailored to improve durability and reduce damages, currently installed components need methods to reduce failures during their exploitation. The present work introduce mathematical problems that lie at the intersection between optimal energy collection, and the prevention of REPAS breakdowns in the solar field. To our knowledge, there are no existing studies on the equilibrium between the two aspects. Therefore, our attention is directed towards literature that delves into the optimization of energy collection.

Assuming a perfect tracking, some optimization problems can be defined for maximal energy collection in CSP plants. In Carrizosa et al. (2015), both energy collection and cost of a Solar Power Tower (SPT) plant are optimized using an alternating greedy-based heuristic in which both the location of the heliostats and the design of the solar tower are simultaneously considered. The aiming point of the heliostats can also be optimized using Integer Linear Programming algorithms to achieve optimal energy production while reducing the damage to the components (Ashley et al., 2017). Correcting the position of the mirrors can be achieved with the use of a distributed system of sensors (Meligy et al., 2021). Their solution is implemented in a Linear Fresnel Reflector, a CSP plant without tracking mechanism.

An area where there has been significant research on energy production efficiency is predictive control (Lopez-Alvarez et al., 2018; Ruiz-Moreno et al., 2021). Controlling the temperature values of the fluid has the advantage of being more robust under disturbances such as changes in the solar irradiance level, mirror reflectivity or inlet HTF temperature (Camacho and Berenguel, 2012). In this context, fuzzy logic and optimization based on physical models have been proposed (Cirre et al., 2009).

Evidently, in systems with a tracking mechanism and not faulty components, the energy collection is maximized when the collector and the Sun are aligned. Tracking systems may work in open or closed loop, depending if a feedback information is provided to the controller. In the former case, an algorithm generates the desired angles applied to the tracker using Sun information such as position, orientation, and time (Fernández-Ahumada et al., 2017; Zhu et al., 2020). In the later, loop is closed through a solar sensor (Garrido and Diaz, 2016; Sneineh and Salah, 2019); hence tracking quality is influenced by the quality of the sensor and mechanical issues like friction and backlash (Diaz et al., 2018). Another important aspect for solar trackers is the number of tracking axes. Depending on this, the tracking modes can be classified as dual-axis or single-axis solar tracking. In commercial CSP plant with PTC systems, solar tracking is typically performed in a single axis, as dual-axis trackers are of complicated structure and higher cost (Kong et al., 2020). Besides of solar equations or sensor-based control, some algorithms for single-axis tracking include shadow information to perform adjustments in the position of the collector (Saldivar-Aguilera et al., 2023).

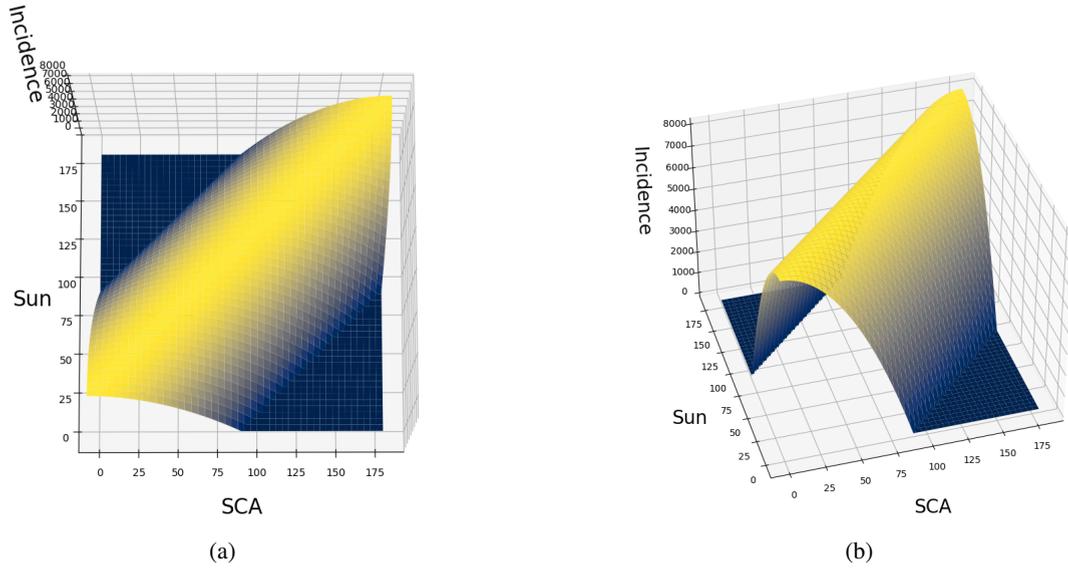

**Figure 2:** Ray incidence over the HCE depending on the angular position of the Sun and the SCA for a simple case. Example assumes that the SCE has a perfect parabolic shape, the HCE is at the focus of the parabola, a tracking system with no errors, and similar operating conditions during the day. (a) x-y perspective, (b) lateral perspective.

The aforementioned works target the problem of perfect tracking, with a special attention in the design of accurate and cost-efficient trackers. In this paper, we study optimization problems providing an alternative solution to existing open loop tracking systems in CSP plants. We consider the rotations of the tracking structure in a single axis. The advantages of existing open-loop trackers includes low cost, no solar sensor requirements, and independence of weather disturbances (Fuentes-Morales et al., 2020). Our main contributions are the definition of novel problems that incorporates the number of movements of the tracking structure in the optimization scheme for CSP plants.

## 3. Problem formulation

Solar irradiance over the HCE can be expressed as a function $z = f(x, y)$, where the $(x, y)$ coordinates represent the SCA and the Sun angular displacements, respectively, while $z$ corresponds to the number of rays touching the HCE. However, if we assume that there is no change in the initial conditions, for instance, a constant weather throughout the day, the 3D surface corresponding to $f$ can be interpreted as a shifted 2D curve. Figure 2 depicts a hypothetical example, when the shifted function shows its maximal amplitude, i.e. higher number of non-zero values at $\frac{\pi}{2}$. This visualization allow us to redefine the function as $z = f(\theta)$, where $\theta$ represents the difference between the Sun and the SCA angular position. Although $f$ can be defined for negatives values, we consider that it is shifted from 0 to a maximum angle $\omega^*$.

Solar tracking is discrete in PTC plants, hence the interval $[0, \omega^*]$ can be divided in $n$ disjoint ordered by $x$ subintervals $S_i = [\theta_{i-1}, \theta_i)$, with $\theta_0 = 0$, such that the number of rays in each of these subintervals is constant, i.e $f(\theta) = \alpha_i$ for each $\theta \in S_i$. Therefore we can describe $f$ as a step function [1] with $n$ steps as follows:

$$f(\theta) = \sum_{i=1}^{n} \alpha_i \delta_i(\theta), \tag{1}$$

where $\alpha_i$ is the number of rays touching the HCE in the interval $S_i$, and $\delta_i$ is the indicator function of the interval $S_i$.

Using a ray-tracing software for CSP plants (Wendelin, 2003; Blanco et al., 2005), $f$ can be obtained by moving the Sun around a fixed axis (North-South) with the SCA at 90º (zenith), as illustrated in Figure 3 for a fixed SCA and Sun position. The events at which the Sun rays start/end intersecting the HCE can occur at any angular difference between the Sun and the SCA position; hence, the length of the steps in $f$ can be any real number. However, in this paper, we consider the case in which these numbers are approximated as rational numbers (e.g., accurate to within one thousandth of a unit). This is standard in real-world applications, and involves the computational representation of real numbers. In the rational case, it is easy to see that the problem can be reduced to one in which the solar irradiance function has steps with integer length.

---

[1]In mathematics, a step function is a piecewise constant function having only finitely many pieces.

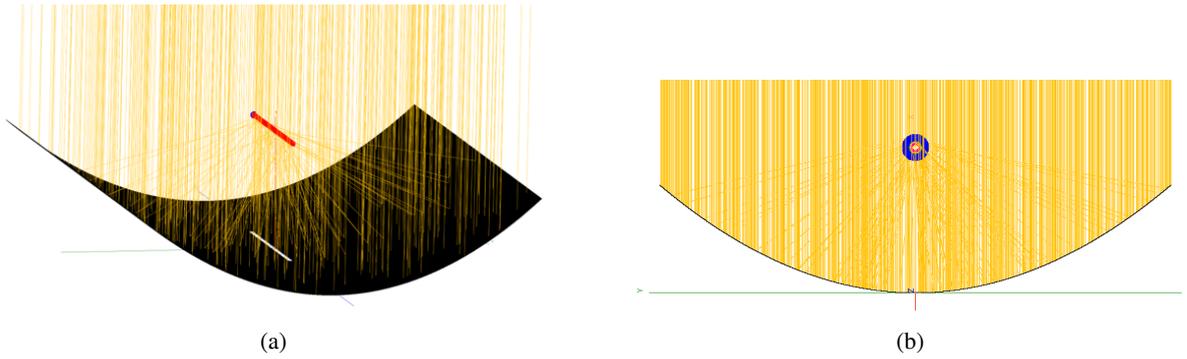

**Figure 3:** Ray tracing representation for a fixed SCA and Sun position: (a) 3 dimensional representation, (b) 2 dimensional cut.

Let $f \colon \mathbb{R} \to \mathbb{N}$ be a step function with $n$ steps, defined as in equation 1. Let $S = \{S_1, \cdots, S_n\}$ be the set of intervals of $f$. As we mentioned, each interval is of the form $S_i = [\theta_{i-1}, \theta_i)$, and in each of these intervals occurs a step of the function $f$. We will call the *edges* of each step as the angles $\theta_{i-1}$ and $\theta_i$ associated to each $S_i$. The set of edges of $f$ will be noted by $E$. Given an interval $t = [a, b]$, we define its *length* as $l_t = b - a$ and its *gain* $g_t$ as the area below the function $f$ in the interval $t$. Mathematically, it is the following integral:

$$g_t = \int_a^b f(\theta) d\theta \tag{2}$$

Clearly, $g_t$ can be computed in $O(n)$ time, where $n$ is the number of steps of $f$.

For a multiset of intervals $T = \{t_1, \dots, t_k\}$, the total solar irradiance (gain) of the set is $I_T = \sum g_{t_i}$, and the total length is defined as $L_T = \sum l_{t_i}$. We will denote by $\tilde{\theta}_s$ to the *initial position of the SCA w.r.t the Sun.* See Figure 4 for an overview of the described notations.

We formulate two optimization problems of particular interest for solar tracking in CSP plants:

**Problem 1.** *(Minimum Tracking Motion, or MTM-Problem): Given a step function $f$ defined on $[0, \omega^*]$, and two real numbers $u_1$, $u_2$; find a set of intervals $T^* = \{t_1, \dots, t_m\}$ of minimum size s.t. $t_i \subseteq [0, \omega^*]$, $\forall \theta \in t_i$, $u_1 \le f(\theta) \le u_2$ and $L_{T^*} + \tilde{\theta}_s = \omega^*$.*

**Problem 2.** *(Maximal Energy Collection, or MEC-Problem): Given a step function $f$ defined on $[0, \omega^*]$ and $m \in \mathbb{N}$, find a set of intervals $T^* = \{t_1, \dots, t_j\}$ s.t. $t_i \subseteq [0, \omega^*]$, $|T^*| \le m$, $L_{T^*} \le \omega^*$, and $I_{T^*}$ is maximal.*

Problem 1 translates to finding the minimum number of movements of the SCA such that the solar irradiance intersecting the HCE at any moment is preserved within a given range. On the other hand, Problem 2 addresses to optimize the total solar irradiance intersecting the HCE. In this case, the constraints are defined as the number of movements allowed to be performed on the SCA, and the maximum displacement the SCA can cover. For simplicity, when describing the solutions of the aforementioned problems, we use the term set for referring to multiset of intervals.

## 4. Minimum Tracking Motion

The analysis of the MTM-Problem necessitates consideration of the initial positioning of the SCA relative to the Sun position. Two cases are possible: one where the SCA resides in a feasible configuration, i.e. $u_1 \le f(\tilde{\theta}_s) \le u_2$; and another where the SCA violates this constraint. In the former scenario, it is evident that the optimal solution involves waiting until the Sun transitions to a non-feasible state is reached. Consequently, we can assume, without loss of generality, that the SCA starts from a non-feasible configuration.

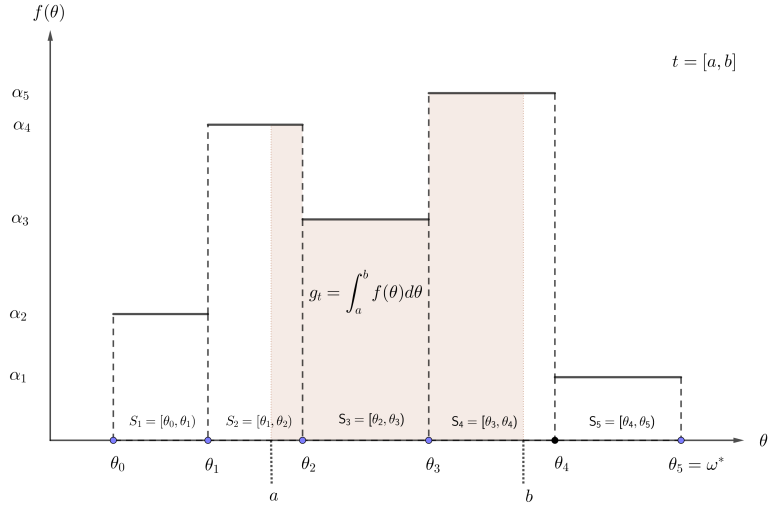

**Figure 4:** Main elements defining the ray incidence function $f$. $t$ is a tracking interval from $a$ to $b$. Total irradiance in $t$ ($g_t$) is the area within $t$ below the histogram. Although $f$ can be defined for negative values, we consider it shifted to the range $(0, \omega^*)$.

**Theorem 1.** *Let $t^* = (\theta_i, \theta_j)$ be a maximum size tracking interval in $f$ such that for any $\theta \in t^*$, $u_1 \leq f(\theta) \leq u_2$. If the SCA initially violates the boundary conditions, then the minimum cardinality of rotations required for a solution to satisfy the conditions of the MTM-Problem is $\lceil \frac{\omega^* - \tilde{\theta}_s}{l_{t^*}} \rceil$ .*

*Proof.* Since the SCA violates the boundary condition, it needs to be moved to a feasible configuration. Let us assume that such feasible configuration initiates at $\theta_i$ and when the Sun reaches $\theta_j$ the SCA moves again to $\theta_i$. In such case, it is clear that the SCA has rotated $m = \lceil \frac{\omega^* - \tilde{\theta}_s}{l_{t^*}} \rceil$ times. Hence $T^* = \{t^*, \ldots, t^*, \hat{t}\}$ with the size of $T^*$ equals $m$ is a feasible solution of Problem 1, where $\hat{t} \subseteq t^*$ has length so that $L_{T^*} = \omega^* - \tilde{\theta}_s$. Moreover, $T^*$ is of miniminum size; otherwise, if $T = \{t_1, \ldots, t_r\}$ is a feasible solution with $r < m$ and $L_T = L_{T^*}$, then there would exist a $t_i$ whose length is larger than the length of $t^*$, this contradicts the maximality of $t^*$. $\square$

**Corollary 2.** *The MTM-Problem can be solved in $O(n+m)$ time, where $n$ is the number of steps in $f$ and $m$ is the size of a solution.*

*Proof.* The proof of Theorem 1 provides an additional insight on the optimal value when the SCA starts from a feasible configuration. If $l_0$ is the length of the interval in which the SCA meets the problem restrictions from the begining, then the minimum number of rotations of the SCA is $m = \lceil \frac{\omega^* - \tilde{\theta}_s - l_0}{l_{t^*}} \rceil$. Finally, since the maximal interval $t^*$ can be computed in linear time with a sweep from left to right, a greedy algorithm computes the optimal solution $T^*$ in $O(n+m)$ time. $\square$

**Remark 1.** *MTM-Problem can be solved in $O(n)$ with a multiset representation.*

## 5. Maximal energy collection

We say that a solar irradiance function $f$ is *unimodal* if, for exactly one $i \in \{1, \ldots, n\}$, it is satisfied that $f(\theta_{j-1}) \leq f(\theta_j)$, $\forall j \leq i$ and $f(\theta_j) \geq f(\theta_{j+1})$, $\forall j \geq i$, where $\theta_{j-1}$, $\theta_j$ and $\theta_{j+1}$ are edges of $f$. In such case, we will say that the interval $S_i$ is its *maximum*. Analogously, we can define an interval $S_i$ to be *a local maximum* if $f(\theta_{i-1}) \leq f(\theta_i)$ and $f(\theta_i) \geq f(\theta_{i+1})$. We say that $f$ is multimodal or *k-modal* if it has $k$ local maxima. In this section, we propose different approaches to solve MEC-Problem depending on the modality of $f$.

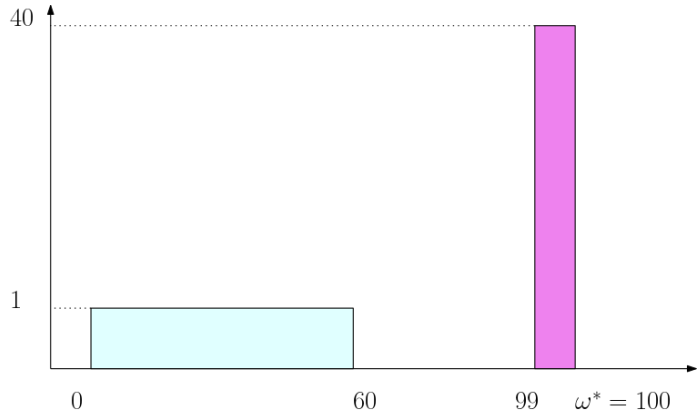

**Figure 5:** Example in which the greedy approach does not work for $m = 3$ and $k = 2$. The greedy algorithm for $m = 3$ and $\omega^* = 100$ centered at the left local maximum (the blue one) gives a total gain of 100, while if it is centered at the right local maximum (the violet one), the total gain would be 120. However, if we took two intervals centered at the right of length 1 (the violet rectangle two times) and one at the left of maximum possible length, we would get a gain of 140.

### 5.1. Unimodal

Given a real number $l$, let $G_l$ be the maximum gain with respect to $f$ of an interval of length $l$. By simplicity, we refer to $G_l$ as the *maximum gain of length $l$*. Notice that when $f$ is unimodal, any interval of maximum gain of length $l$ contains intervals of maximum gain for lengths lower than $l$. Hence, given $l_1, l_2 \in \mathbb{R}$ with $l_1 \le l_2 \le \omega^*$, it can always be found intervals $t_1$ and $t_2$ of lengths $l_{t_1} = l_1$ and $l_{t_2} = l_2$ of maximum gain in $f$ for $l_1, l_2$, respectively, such that $t_1 \subseteq t_2$. In this case, we say that $t_1$ is *embedded* in $t_2$.

**Lemma 3.** *Let $l_1, l_2$ be two numbers s.t. $l_1 < l_2$. Then $G_{l_1} + G_{l_2} \le G_{l_1 + \epsilon} + G_{l_2 - \epsilon}$, for all $0 \le \epsilon \le l_2 - l_1$.*

*Proof.* Let $t_1 = [a_1, b_1]$ and $t_2 = [a_2, b_2]$ be two maximum gain intervals associated to $l_1, l_2$ respectively such that $t_1 \subseteq t_2$. Observe that, given the unimodal nature of $f$, in the interval $[b_1, b_2]$ the function is non-increasing, while in the interval $[a_2, a_1]$ it is non-decreasing. Then, for every $\epsilon \le b_2 - b_1$ (alternatively, $\epsilon \le a_1 - a_2$), the intervals $t_1' = [a_1, b_1 + \epsilon]$ and $t_2' = [a_2, b_2 - \epsilon]$ (alternatively $t_1' = [a_1 - \epsilon, b_1]$ and $t_2' = [a_2 + \epsilon, b_2 - \epsilon]$) satisfy $G_{l_1} + G_{l_2} \le g_{t_1'} + g_{t_2'} \le G_{l_1 + \epsilon} + G_{l_2 - \epsilon}$. Since the intervals $[a_2, a_1]$ and $[b_1, b_2]$ are disjoint, then the Lemma holds for all $0 \le \epsilon \le a_1 - a_2 + b_2 - b_1 = l_2 - l_1$. $\square$

**Theorem 4.** *Let $l = \frac{\omega^*}{m}$ and $t$ be a subinterval of $[0, \omega^*]$ s.t. $l_t = l$ and $g_t = G_l$. Then, $T^* = \{t, ..., t\}$ with $|T^*| = m$ is optimal for MEC-Problem when $f$ is unimodal.*

*Proof.* The solution is feasible since $|T^*| \le m$ and $L_{T^*} = \omega^* \le \omega^*$. Let us consider an optimal set of intervals $T = \{t_1, ..., t_m\}$. Let $t_i \in T$ s.t. $l_{t_i} > l_t$, then there exists $t_j \in T$ s.t. $l_{t_j} < l_t$; hence $t_j$ can be embedded into $t_i$. Therefore, by Lemma 3, $t_j$ can be increased while reducing $t_i$ in a way such that the total length remains $\omega^*$ but with no less total incidence; in particular, for $\epsilon = l_t - l_j \le l_i - l_j$ we obtain an interval $t_j'$ from $t_j$ with length exactly $l_t$. We follow this procedure until only two intervals $t_i$ and $t_j$ have length different to $l_t$. In this last step, after applying the procedure, all intervals have length exactly $l_t$ because $L_T = \omega^*$. Finally, $T' = \{t_1', ..., t_m'\}$ with $l_{t_j'} = l_t$, for $j = 1, ..., m$, can be replaced by $T^*$ since $t$ is an interval of maximum gain with length $l$. $\square$

**Remark 2.** *Notice that the previous results are valid for any unimodal function $f$, not necessarily with steps of integer length.*

Using Theorem 4, a greedy strategy solves the problem and we arrive to the following result:

**Corollary 5.** *The MEC-Problem can be solved in $O(n)$ time when $f$ is unimodal.*

## 5.2. General case

Firstly, it is important to note that the proposed greedy algorithm does not work when $f$ is $k$-modal ($k > 1$). See, for instance, the example showed in Figure 5.

Let us introduce a some concepts that will be useful in our approach to solving the problem when the function $f$ has $k > 1$ local maxima. Recall that $E$ is defined as the set of edges of $f$.

**Definition 1.** *An interval $t = (\theta_1, \theta_2)$ is discrete, called as a d-interval, if both $\theta_1$ and $\theta_2$ belong to $E$. The interval is semi-discrete if it starts or ends in an edge of $f$.*

**Definition 2.** *An interval $t = (\theta_1, \theta_2)$ is an md-interval, if it is discrete and contains at least a local maximum. A semi md-interval is a semi-discrete interval containing at least a local maximum of $f$.*

### 5.2.1. Properties

**Lemma 6.** *There exists an optimal solution $T^*$ to MEC-Problem such that, for any $t \in T^*$, $t$ is semi-discrete.*

*Proof.* Let $T'$ be an optimal solution to MEC-Problem, and $t = (\theta_1, \theta_2)$ be any interval belonging to $T^*$, with $\theta_1$ and $\theta_2$ not necessarily in $E$; then either $f(\theta_1) = f(\theta_2)$ or $f(\theta_1) \neq f(\theta_2)$. In the former case, we can move $t$ to the left/right until the closer edge is reached, obtaining a semi-discrete interval $t'$. On the later, without loss of generality, assume $f(\theta_1) > f(\theta_2)$. Then $\theta_1 \in E$, otherwise we can move $t$ to the left, increasing the optimal value; hence $t$ is semi-discrete. Therefore, we can transform $T'$ into a solution $T^*$ where all intervals are semi-discrete. □

**Theorem 7.** *There exists an optimal solution $T^*$ to MEC-Problem such that, $\forall i = 1 \ldots |T^*| - 1$, $t_i$ is a discrete interval.*

*Proof.* Let $T'$ be an optimal solution to the MEC problem s.t. all intervals are at least semi discrete (see Lemma 6). Let $t_1 \in T'$, $t_2 \in T'$ be two semi discrete intervals and, $\theta_1, \theta_2$ be two extremes of $t_1, t_2$, respectively, that are not edges of $f$. Notice $f(\theta_1) = f(\theta_2)$, otherwise $T'$ is not optimal; then we can increase $t_1$ and reduce $t_2$ until one of the edges of $f$ is reached. This transformation can be applied to every pair of semi discrete intervals in $T'$; hence $T^*$ can be obtained with $|T^*| - 1$ $d$-intervals. As order is not important in any optimal solution to the MEC problem, $T^*$ can be arranged s.t. the interval being at least semi discrete is the last one. □

Since $|E| = n + 1$, the number of $d$-intervals that can be formed from the steps of $f$ is in $O(n^2)$. However, not all of these intervals are candidates to an optimal solution. Notice that solutions not including local maximums are not of maximal gain. Then, we have:

**Lemma 8.** *Let $t$ be an interval of the optimal solution to MEC-Problem. Then $t$ contained within an md-interval.*

Joining the findings of Lemma 8 and Theorem 7, we establish the main result of this section:

**Theorem 9.** *There exists an optimal solution $T^*$ to MEC-Problem such that, $\forall i = 1 \ldots |T^*| - 1$, $t_i$ is an md-interval, and $t_m$ is a semi md-interval.*

Theorem 9 does not reduce the number of intervals analyzed in terms of the worst case computing-time. However, it provides a nice characterization of an optimal solution to MEC-Problem. In addition, as we will show in the next section, the outcomes obtained in this section enable the efficient computation of a solution by decomposing the optimization problem into two distinct tasks.

### 5.2.2. The approach

The following property of any optimal solution $T^*$ can be easily verified: *removing any interval $t_i$ from $T^*$ yields a solution $T' = T^* - \{t_i\}$ which is optimal for $m - 1$ moves and $\omega^* - l_{t_i}$ total displacement of the SCA.* This observation ensures that the problem possesses the *optimal substructure property*, which, in essence, permits the derivation of an optimal solution by solving a set of interconnected subproblems. In addition, and more importantly, according to Theorem 9, the general form of the optimal solution to MEC-Problem can be expressed as:

$$I_m^* = D_{m-1}^l + G_{\omega^* - l}, \tag{3}$$

being $I_m^*$ the maximum gain associated to $m$ moves, $D_{m-1}^l$ the maximum gain for length at most $l$ using $m - 1$ discrete intervals, and $G_{\omega^* - l}$ the maximum gain in $f$ for the remaining length. Because of the optimal substructure of the problem, $D_{m-1}^l$ is optimal for length $l$. However, we cannot know beforehand the value of $l$, hence we divide the problem in two tasks:

Task 1: Compute the optimal value $D_{m-1}^l$, $\forall l \in (0, \omega^*)$.

Task 2: Compute $G_l$, $\forall l \in (0, \omega^*)$.

According to (3), a solution with length $l$ for the first task is associated to a solution with length $\omega^* - l'$ in the second, where $l' \leq l$ is the total length of the intervals obtained during the computation of $D_{m-1}^l$. In addition, notice that $l \in \mathbb{N}$ since the length of the steps of $f$ are integers. Therefore, the following remarks are straightforward:

**Remark 3.** *Combining the solutions from Task 1 and Task 2 takes* $O(\omega^*)$.

**Remark 4.** $I_m^*$ *is the maximum value obtained after combining the solutions from Task 1 and Task 2.*

**Theorem 10.** *Task 2 can be solved in* $O(n\omega^*)$ *time.*

*Proof.* By Theorem 9, the last interval in the optimal solution is a semi md-interval, hence is semi discrete by Definition 2. Using Definition 1, $G_l$ can be obtained for a given $l$ by checking every edge $e$ of $f$ twice, assuming the interval starts/ends in $e$. As $|E| = n + 1$ and $l \in \mathbb{N}$, a list with $G_l$ values can be obtained in $O(n\omega^*)$. $\qquad\square$

We now focus on solving Task 1. Since the considered intervals are discrete for this task, we are able to design an efficient algorithm based on dynamic programming (DP). Our algorithm will solve MEC-Problem for any length considering only *md*-intervals, which is the requirement for Task 1. For simplicity, we refer to this version as the discrete-MEC-Problem.

Let $B$ be the set containing the *md*-intervals of $f$. In addition, let us consider the table $D[i, j, l]$ indicating the maximum gain for discrete-MEC-Problem when using up to interval $i$ of $B$, $b_i$, with $j$ movements and $l$ as maximum solar displacement. Notice that intervals in $B$ do not need to be sorted, but we assume a fixed order (labelling) during the execution of the algorithm. Thus, the recursive equation for $D$ can be expressed as:

$$D[i, j, l] = \begin{cases} 0 & \text{(a) } 0 \in \{i, j, l\} \\ D[i-1, j, l] & \text{(b) } l < l_i \\ \max(D[i-1, j, l], g_i + D[i, j-1, l-l_i]) & \text{(c) } else, \end{cases} \quad (4)$$

where $g_i$ represents the gain of the interval $i$ of $B$.

**Theorem 11.** $D$ *gives an optimal solution for discrete-MEC-Problem.*

*Proof.* Let us see that $D[i, j, l]$ is optimal for any value of $i, j, l$. According to equation (4), we divide the proof by cases. Let $b_i$ be the interval with index $i$ in $B$.

(a) If any variable is 0, then there is no associated gain. This is a base case, and not expected in a real situation.

(b) In this case, $(l < l_i)$, $b_i$ is an interval of length bigger than the target length. Since intervals must be used in their total extension, $b_i$ is not part of the optimal solution.

(c) Assume that $j > 0$ and $l \geq l_i$. In any solution, $b_i$ is either used or not. On the one hand, the optimal solution when $b_i$ is not used is stored in $D[i-1, j, l]$. On the other hand, the best value when using $b_i$ is given by $g_{b_i} + D[i, j-1, l-l_i]$. Therefore, the optimal solution using up to interval $i$ is the maximum of the aforementioned values. $\qquad\square$

Following Theorem 11, we emphasize how to solve Task 1 for any target length:

**Remark 5.** *For a given* $l \in (0, \omega^*) \cap \mathbb{N}$, $D[|B|, m-1, l]$ *contains the optimal value for Task 1.*

**Theorem 12.** $D$ *can be computed in* $O(n^2 m\omega^*)$ *time.*

*Proof.* At each time step, depending on the evaluated condition (a), (b), or (c), there are at most two queries to values stored in $D$, which are solved in constant time (including $g_i$, which can be precomputed for all the intervals). There are at most $n^2$ md-intervals in $B$, $m$ movements, and $\omega^*$ solar displacement; hence, the required running time is $O(n^2 m\omega^*)$. $\qquad\square$

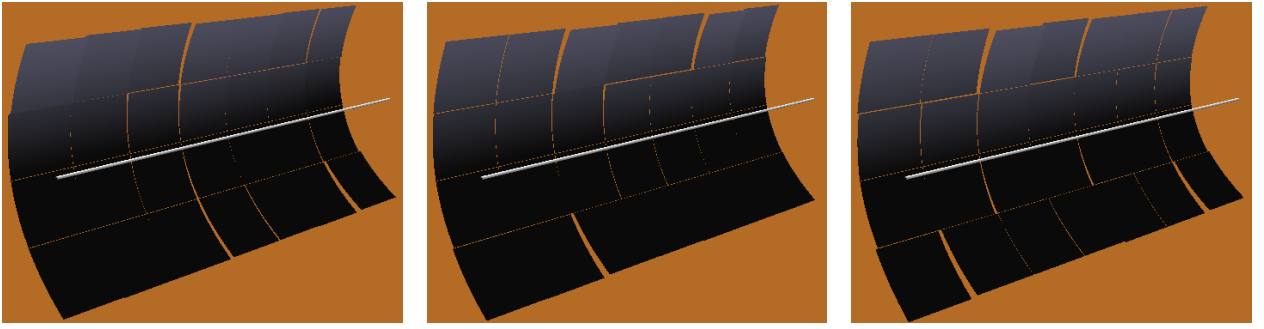

**Figure 6:** Three baseline cases for experimentation. The angle of every mirror is perturbed randomly.

The intervals corresponding to an optimal solution $T^*$ to the MEC-Problem can be obtained after computing $I_m^*$. Notice that every decision is associated to an interval, both in Task 1 and Task 2; see Theorems 10 and 11. In Task 2, the interval associated to $G_l$ (for a given value of $l$) can be obtained by a sweeping approach on $f$. On the other hand, when addressing Task 1, it is more convenient to utilize the cases defining equation (4) for retrieving the intervals associated with a decision. Specifically, for any $i, j, l$, we check (a), (b) or (c); if cases (a) or (b) holds, then the candidate $i$ is not used; if $c$ holds, then we check the equality $D[i, j, l] = D[i-1, j, l]$ and if it holds, then candidate $i$ is not used, otherwise, it is used. Starting this process at the cell $D[|B|, m-1, l]$, being $l$ the length of the optimal solution, it is feasible to extract the complete set of intervals in $T^*$, leading to the following result:

**Corollary 13.** *MEC-Problem can be solved in $O(n^2 m \omega^*)$.*

*Proof.* This follows directly from Theorem 12, Theorem 10, and Remark 3, being the total running time $T(n) = O(n^2 m \omega^*) + O(n \omega^*) + O(\omega^*)$; hence $T(n) = O(n^2 m \omega^*)$. □

## 6. Experiments

Two sets of experiments are presented to analyse the proposed solution of MEC-Problem in the $k$-modal case. First, we analyse SCEs failures individually and observe their impact on the shape of $f$; then 12 SCEs are combined to perform the study over the entire SCA. The experiments are conducted with the ray tracing software Tonatiuh (Blanco et al., 2005).

### 6.1. The Solar Collector Element (SCE)

We design the SCE as a parabolic surface divided in 7x4 mirrors. The aperture of the parabola is 5.6m, and the total length of the SCE is 12m. Although the SCE is composed of three (Heat Collector Elements) HCEs, in our simulation we join them in one structure of 12m length and 3.5cm of radius. To account for operational stress in the supporting structure, we assume every mirror is rotated at least by 0.2º around an axis parallel to the tracking axis. Furthermore, we add random perturbations to the angle of each mirror, sampled from a uniform distribution in the range (-2º, 2º), and obtain three baseline cases for experimentation; see Figure 6. We consider three plausible scenarios in PTC plants:

1. No additional failures.

2. The HCE is rotated perpendicular to the tracking axis by 1º.

3. The HCE is broken (simulated with a smaller radius), and three mirrors are missing.

We compute the $f$ function for each of the aforementioned scenarios, by placing the SCE at 90º and rotating the Sun above it. The Sun is rotated in 1º intervals, hence $\omega^* = 180$. In total, we consider 9 functions, as depicted in Figure 7, accounting for the three baseline cases in each of the predefined scenarios. We compare our proposed solution for the multi-modal case (termed as DP) with two additional algorithms:

- **CSP-Track.** Shifting the collector by increments of 1º, up to the maximum angle. Since the Sun displacements are performed in this angle, this is assumed to be optimal, hence we emulate the current trackers in PTC plants.

| Algorithm | Total rays | Percentage decrease | | |
|---|---|---|---|---|
| | $1 * m$ | $.75 * m$ | $.5 * m$ | $.25 * m$ |
| Greedy | 7.53e7 | -0.27% | -0.08% | -0.32% |
| DP (ours) | 7.53e7 | -0.04% | -0.08% | -0.25% |
| CSP-Track | 7.53e7 | - | | |

(a) **Scenario:** No additional failures.

| Algorithm | Total rays | Percentage decrease | | |
|---|---|---|---|---|
| | $1 * m$ | $.75 * m$ | $.5 * m$ | $.25 * m$ |
| Greedy | 5.26e7 | -0.27% | -0.11% | -0.28% |
| DP (ours) | 5.26e7 | -0.06% | -0.11% | -0.20% |
| CSP-Track | 5.26e7 | - | | |

(b) **Scenario:** HCE rotated.

| Algorithm | Total rays | Percentage decrease | | |
|---|---|---|---|---|
| | $1 * m$ | $.75 * m$ | $.5 * m$ | $.25 * m$ |
| Greedy | 8.2e7 | -0.27% | -0.16% | -0.44% |
| DP (ours) | 8.2e7 | -0.09% | -0.16% | -0.39% |
| CSP-Track | 8.2e7 | - | | |

(c) **Scenario:** Broken mirrors.

**Table 1**
Decrease in percentage on the number of rays intersecting the HCE, depending on the number of moves ($m = 26$) used in each algorithm. The results are averaged in the three random cases specified in Figure 6.

- **Greedy.** The greedy algorithm for the unimodal case. This algorithm has the same behaviour as CSP-Track when the collector is moved at a fixed angle, not necessarily 1º.

Recall that our objective is to investigate methods for minimizing the movements of the Solar Collector Equipment (SCE) while maximizing the collection of substantial energy. To this end, we do not consider the full range of $f$. As depicted in Figure 7, all functions have a central portion where the rays are more concentrated; hence we analyze our algorithm by cropping $f$ in the range of (75º, 101º), see Figure 8. Ray concentration in one area of $f$ can be easily explained by the small acceptance range of PTC systems; recall the aperture width of the parabola is 5.6m while the radius of the HCE tube is only 3.5cm. Therefore, the vast majority of the rays strike the small tube in a limited range.

Tables 1a, 1b, 1c depict the production decrease in percentage when reducing the number of movements according to the three considered scenarios respectively: *no additional failures*, *HCE rotated*, and *broken mirros and HCE*. In each scenario, the three baseline cases presented in Figure 6 are evaluated and their solutions are averaged. Notice that percentage decrease is not included for CSP-Track because this algorithm does not consider the number of rotations of the SCA. The proposed Greedy algorithm is expected to simulate the behaviour of CSP-Track for a given value of $m$. However, we retain the use of CSP-Track as a direct indicator of the optimal solution for the maximum value of $m$. We will use this strategy in all the presented results.

The analysis of Tables 1a, 1b, 1c allows to draw some interesting conclusions. In all the considered scenarios, when using 75% of the total movements, the number of rays intersecting the HCE surpass the 90% of the initial rays for DP solution; thus the energy efficiency is assured with a considerable reduction in the number of rotations. In addition, Greedy solution is close to the optimal solution when using half of the initial movements. This can be explained by realizing that the central region of $f$ retains the higher number of rays. Finally, it can be observed that the total number of rays in Table 1c is higher than the total number of rays in Table 1a, where no additional failures are considered. This is due to a fixed number of rays intersecting the collector in each simulation, which depends on the shape of the collector. In other words, with three missing mirrors (Table 1c), the software readjust the ray casting algorithm so that the number of rays intersecting the SCE is the same that when the three mirrors are in place. Therefore, the number of rays intersecting the SCE is constant in all simulations, and what changes is the number of rays intersecting the HCE, which is independent of the number of mirrors.

Although it can be interesting to observe the evolution of the gain as a function of $m$, we reserve this result for a more complex scenario: tracking the Sun for the full SCA.

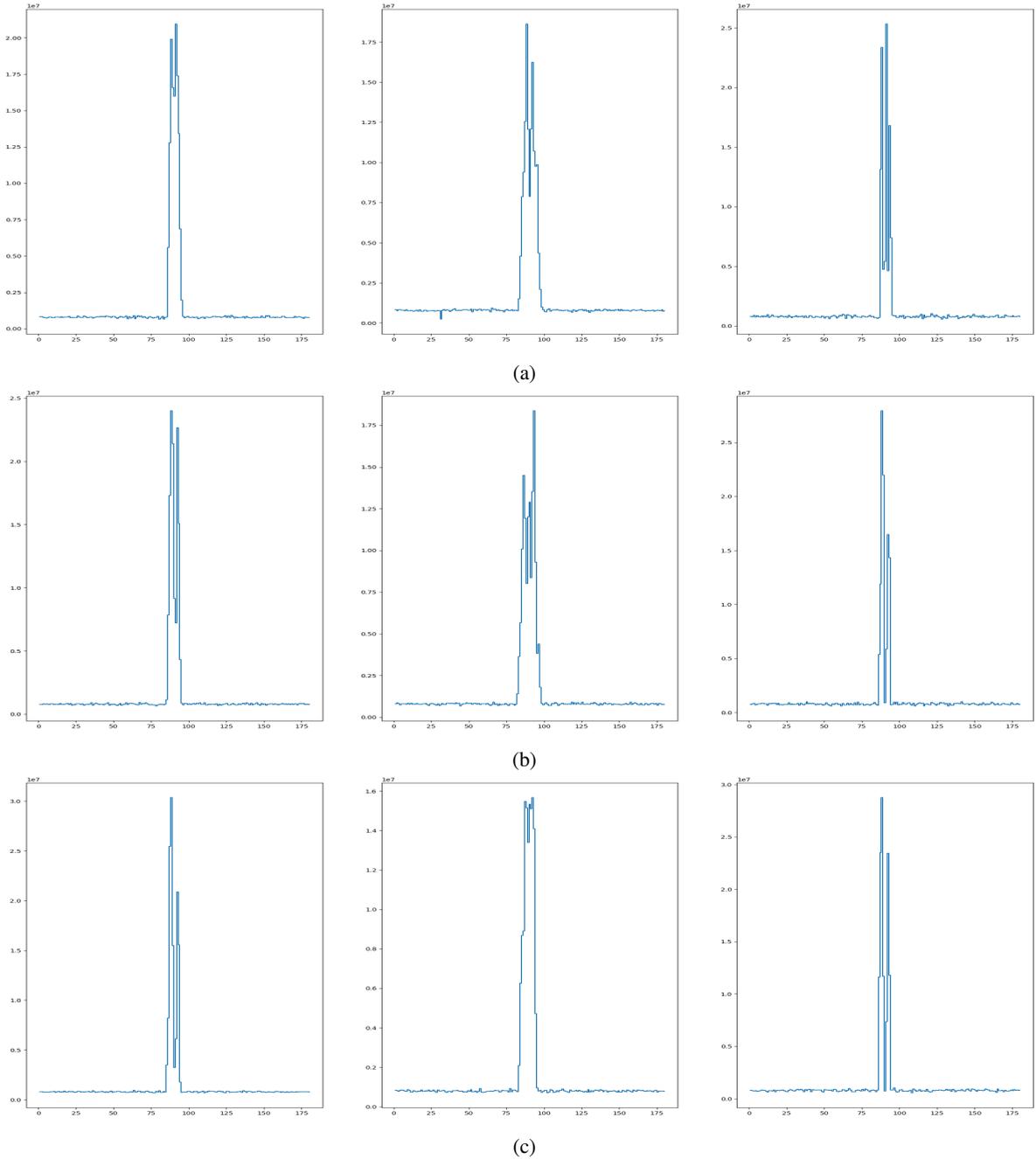

**Figure 7:** Function $f$ for 9 different scenarios. **x-axis**: angular difference between the Sun and the SCE position., **y-axis**: number of rays intersecting the HCE (incidence). In each row the angle of the mirrors in the SCE have the same perturbation, as indicated in the three baseline cases of Figure 6. From left to right, each column correspond to: 1) no additional failures, 2) HCE rotated vertically 1º, and 3) broken HCE glass cover and three broken mirrors.

## 6.2. The Solar Collector Assembly (SCA)

Recall that an SCA in a typical PTC plant is composed of 12 SCEs. To simulate possible failures in such a system, we select 12 times a random function from the ones depicted in Figure 7, and each function is assigned to an SCE. Since consecutive SCEs may have a rotation displacement; i.e., one is at 90º and the other at 91º, we apply a random integer shift to $f$ in the range (-3º, 3º). Notice that rotating the SCE by $d$ degrees is similar to shifting $f$ by the same amount. The final $f$ for the SCA is obtained with the sum of the 12 functions of the SCEs. To simulate a full loop in a PTC plant, we apply the described procedure four times. As in the case of the SCE, when computing $f$, we crop it to the interval (75°, 101°), hence $\omega^* = 26$.

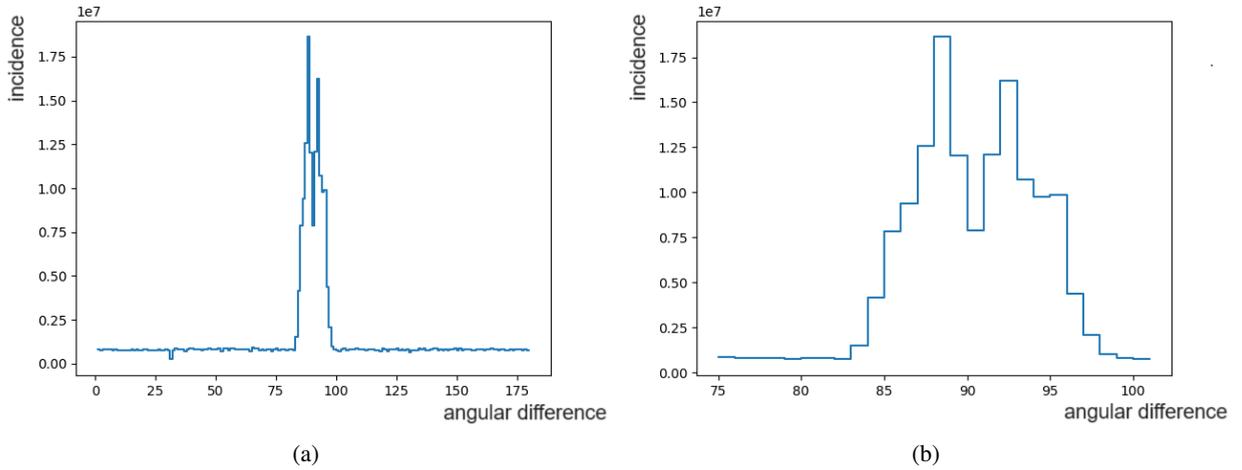

**Figure 8:** Example of the region of interest. (a) the full simulated function, (b) central portion, where the rays are more concentrated. $y$ axis represent the ray incidence over the HCE; $x$ axis is the angular difference between the Sun and the SCE position.

In this experiment, the result is presented as an evolution of the ray incidence over the HCEs according to the number of movements performed by the SCA while tracking the Sun in the four random scenarios; see Figure 9. Notice the importance of DP when the goal is to reduce the number of rotations of the SCA while maintaining a high performance. Unlike the Greedy algorithm, DP increases the performance when more movements are available. In fact, the shape of the function obtained with the Greedy algorithm indicates a similar behaviour in the considered scenarios, thus, it is not able of adapting to the shape of $f$. Conversely, when employing Dynamic Programming (DP), the optimal instances for reducing the number of movements become evident, along with the methodological approach to do so. It can be observed cases when reducing more than half of the movements results in a small decrement of the ray incidence function, which confirms the interest of the algorithm when a large number of moves must be avoided.

## 7. On the NP-Hardness of MEC-Problem

At a first glance, the MEC-Problem introduced in this paper seems hard. However, we have proved that the problem is polynomial in $n$ when the steps in the ray incidence function $f$ have integer length. In this section, we analyze the implications when relaxing the mentioned restriction.

We define some auxiliary notations. Let $S_Z$ be the set of steps of $f$ with integer length; $S$ stands for the general case; let $I_D$ be the set of discrete intervals, i.e. their ending points correspond to edges of steps; finally, let $I$ be the set of intervals in the general case. Subsequently, some variants can be defined using tuples as follows:

1. $(S, I_D)$: MEC-Problem when the steps are of any length, and only discrete intervals are considered in the optimal solution.

2. $(S_Z, I)$: MEC-Problem when the steps are of integer length, without restrictions in the intervals of the optimal solution.

3. $(S, I)$: MEC-Problem when the steps are of any length, without restrictions in the intervals of the optimal solution.

First, notice that the solution provided for the MTM-Problem, as well as the unimodal case of the MEC-Problem, works for steps of any length. Furthermore, the characterization of the optimal solution in Theorem 9 is also extensible to any case, in particular for the general case $(S, I)$. In the following, we prove that $(S, I_D)$ is NP-Hard. Interesting, $(S_Z, I)$, the MEC-Problem addressed in this paper, can be solved with a pseudo-polynomial algorithm (polynomial in $|S_Z|$)[2]. Notice that the algorithm presented in this research includes $(S_Z, I_D)$, hence this variant can also be solved in the same computational time. Finally, we conjecture that the more general variant, $(S, I)$, is NP-Hard.

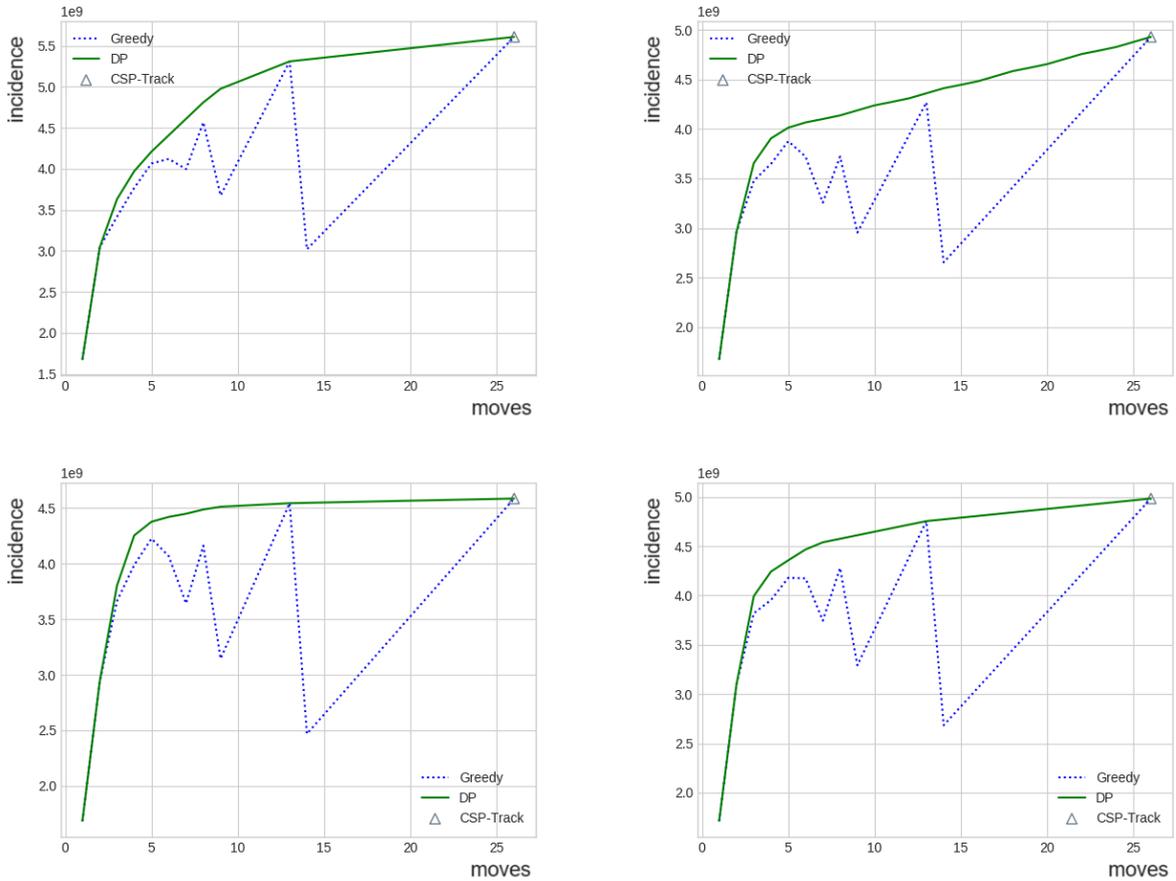

**Figure 9:** Evolution of the ray incidence over the HCE when the SCA is tracking the Sun according to the considered algorithms. CSP-Track emulates the current solution implemented in CSP plants, moving optimally at every time step (1º in the considered scenarios). For this tracking approach, $m = 26$. The $x$ axis represents the number of rotations performed by the SCA, while $y$ axis is the number of rays intersecting the HCEs in the SCA. Four random but realistic scenarios are considered.

In these formulations we do not force $\omega^*$ to be the sum of the steps in $f$, but it is instead an input of the problem. Contemplating a reduced value for $\omega^*$ holds practical implications. For instance, in scenarios involving a cleaning period for the SCE, a shorter tracking time is required.

---

[2]By Corollary 13, the computational time is $O(n^2 m \omega^*)$. The value $\omega^*$, the total length of $S_Z$, is exponential in the number of bits needed for its representation. However, from a practical perspective, $\omega^*$ can be considered constant.

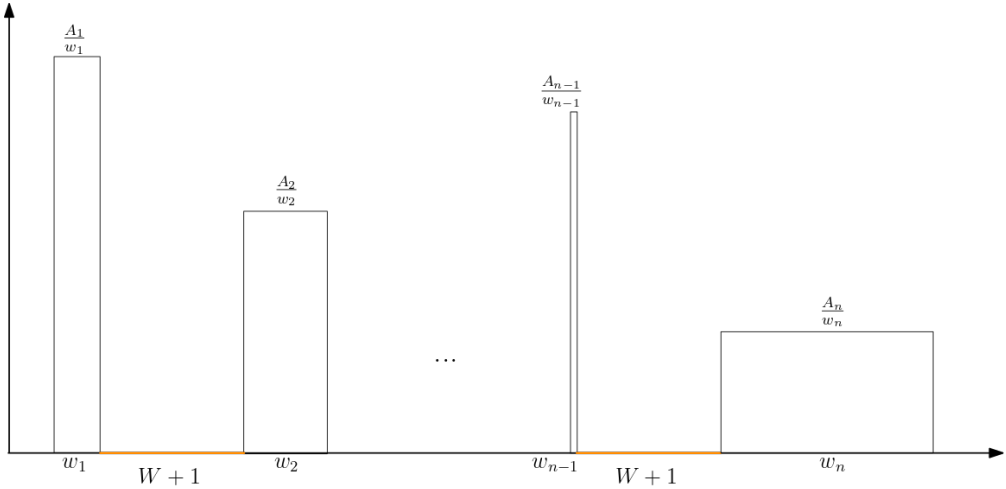

**Figure 10:** Definition of $f$ from an instance of the unbounded Knapsack Problem with $n$ items. $w_j, A_j$ are the weight and profit, respectively, of the item $j$, and $W$ is the capacity of the knapsack.

## 7.1. Knapsack reduction

We show that $(S, I_D)$ is NP-Hard by a reduction from the unbounded Knapsack Problem (u-KP), which has the following formulation:

$$\max \sum_{j=1}^{n} x_j A_j$$

$$\text{s.t. } \sum_{j=1}^{n} x_j w_j \leq W, \tag{5}$$

$$x_j \geq 0, x_j \in \mathcal{N}, j \in [1..n],$$

where $W$ is the capacity of the knapsack. An instance of the u-KP is a set of items with capacity $w_j$ and profit $A_j$, where each item can be placed any number of times within the backpack as long as the capacity constraint is met.

Let us prove the reduction. Given an instance of the u-KP, we need to transform it to an instance of our target problem. To that end, we build $f$ by placing item $j$ (in any given ordering) as a step (interval) $S_j$ of length $w_j$ and $\alpha_{S_j} = \frac{A_j}{w_j}$, hence $g_j = A_j$. In addition, between consecutive steps corresponding to items of the knapsack we place an auxiliary step of length $W + 1$ and gain 0; see Figure 10. Clearly, building $f$ from the input items of the u-KP takes linear running time. Additionally, we need to consider the values of $m$ and $\omega^*$. The later is defined as $\omega^* = W$ to ensure that the knapsack capacity is not surpassed. The former puts a limit on the cardinality of the solution. However, despite of its name, the u-KP has a limit on the number of times an item can be placed inside the knapsack, enforced by the capacity constraint. For an item $j$, this limit is $\lfloor \frac{W}{w_j} \rfloor$, hence we take $m = \sum_{j=1}^{n} \lfloor \frac{W}{w_j} \rfloor$. With the considered values of $f$, $m$, and $\omega^*$, an instance of the u-KP is transformed to an instance of MEC-Problem in polynomial time.

A solution of the $(S, I_D)$ problem maximizes the gain associated to the discrete intervals of $f$ that are feasible. Solving the $(S, I_D)$ version of MEC-Problem implies that only the steps associated to items of the knapsack are considered as valid intervals in the optimal solution. Notice that any other discrete interval has bigger length than $\omega^* = W$. Finally, as the value of $m$ does not remove any candidate solution to the u-KP, and the considered intervals precisely correspond to the items of the knapsack with their respective gains, the output of the $(S, I_D)$ problem directly serves as a solution for the u-KP.

## 8. Conclusions

Optimizing solar energy collection while reducing the operational stress in solar plants is an interesting and complex problem. Existing solutions aims at increasing the total production without taking into account the mechanical damage induced to the collectors. In this work, we formalize new optimization problems for solar plants using solar tracking. Specifically, we target two problems inspired on PTC plants: *Min-Tracking-Motion* (MTM) asks what is the minimum number of movements for solar tracking while keeping the production within a given range; *Maximal Energy Collection* (MEC) endeavors to capture the maximum energy within a predetermined budget of movements. We target both problems from a practical perspective, considering the scenario where the objective function $f$ can be defined as a step function with steps of integer length. In this context, we are able to provide efficient algorithms based on the Greedy and Dynamic Programming paradigms. In the general case of MEC-Problem, we characterize the optimal solution for any step function, including steps of any length. Our proposed solutions for MTM- and MEC-Problems runs in $O(n)$ and $O(n^2 m \omega^*)$ respectively, being $n$ the number of steps in $f$, $m$ the number of movements of the solar structure, and $\omega^*$ the maximal amplitude angle that the structure can cover. Our algorithms are generic and can be adapted to schedule solar tracking in other CSP systems.

We leave two open problems as part of the future work. First, we aim at providing a solution without constraining the length of the steps in $f$ for the MEC problem. Our initial findings indicates that this problem can be NP-Hard. In addition, our plan includes extending the proposed solution to a 3D setting, acknowledging that weather conditions play a role in affecting energy collection.

## Acknowledgments

This work is partially supported by grants PID2020-114154RB-I00, TED2021-129182B-I00 and and DIN2020-011317 funded by MCIN/AEI/10.13039/501100011033 and the European Union NextGenerationEU/PRTR.

We express our gratitude to T. Todtenhaupt from Virtualmech S.L. for providing assistance with the ray tracing software.